\documentclass{conm-p-l}

\copyrightinfo{2015}{}

\usepackage{graphicx}
\usepackage{subfigure}

\usepackage{amssymb,amsmath,amsthm,amscd}

\newtheorem{theorem}{Theorem}[section]

\theoremstyle{definition}

\theoremstyle{remark}

\numberwithin{equation}{section}

\newcommand{\vth}{\vartheta}

\newcommand{\ga}{\gamma}

\newcommand{\Dl}{\Delta}
\renewcommand{\th}{\theta}
\newcommand{\ra}{\rightarrow}

\newcommand{\sg}{\sigma}

\newcommand{\pa}{\partial}

\newcommand{\hy}{\hat{y}}

\newcommand{\bQ}{\bar{Q}}

\newcommand{\nid}{\noindent}

\newcommand{\Om}{\Omega}

\newcommand{\W}{{\mathcal W}}

\newcommand{\htau}{\hat{\tau}}

\newcommand{\hvth}{\hat{\vartheta}}

\begin{document}
   
\title[On the so-called rogue waves in the nonlinear Schr\"odinger equation]{On the so-called rogue waves in the nonlinear Schr\"odinger equation}

\author{Y. Charles Li}
\address{Department of Mathematics, University of Missouri, 
Columbia, MO 65211, USA}
\email{liyan@missouri.edu}
\urladdr{http://faculty.missouri.edu/~liyan}

\curraddr{}
\thanks{}

\subjclass{Primary 76, 35}

\date{}

\dedicatory{}

\keywords{Rogue water waves,  homoclinic orbits, Peregrine wave, rough dependence on initial data, finite time blowup.}

\begin{abstract}
The mechanism of a rogue water wave is still unknown. One popular conjecture is that the Peregrine wave solution of the nonlinear Schr\"odinger equation (NLS) provides a mechanism. A Peregrine wave solution can be obtained by taking
the infinite spatial period limit to the homoclinic solutions. In this article, from the perspective of the phase space structure of these homoclinic orbits in the infinite dimensional phase space where the NLS defines a dynamical system, we 
exam the observability of these homoclinic orbits (and their approximations). Our conclusion is that these approximate homoclinic orbits are the most observable solutions,and they should correspond to the most common deep ocean 
waves rather than the rare rogue waves. We also discuss other possibilities for the mechanism of a rogue wave: rough dependence on initial data or finite time blow up. 
\end{abstract}

\maketitle
\tableofcontents

\section{Introduction}

The mystery of rogue water waves started from folklores of mariners centuries ago. Their existence was scientifically confirmed on New Year's day 1995 at the Draupner platform in the North Sea. In oceanography, rogue waves are defined as waves with height more than twice the significant wave height (SWH). SWH is the average of the top third wave heights in a wave record. A rogue wave is often a single 
tall wave that is localized in both space and time, and appears without warning in mid-ocean. The key in theoretical understanding of rogue waves is:
\begin{itemize}
  \item What is the mechanism of a rogue wave?
\end{itemize}  
Once the mechanism of a rogue wave is understood, it will be easier to understand the causes in different oceanic environments, that can lead to the mechanism to be in action. The consequences of rogue waves have been suspected for many ship sinking incidents. Due to their importance in application and theory, rogue waves have been extensively studied, for a sample of references, see \cite{DKM08}  \cite{CS12} \cite{AAS09} \cite{CHA11} \cite{SPS13} \cite{GZT13} 
\cite{Cha09} \cite{KP03} \cite{Ala14}.

\section{Observability of approximate homoclinic orbits under the nonlinear Schr\"odinger dynamics}

Can homoclinic orbits or Peregrine wave solutions be responsible for rogue water waves? This is the interesting question asked by 
many researchers \cite{DT99} \cite{DKM08} \cite{AAS09} \cite{CS12}. Peregrine wave solutions ``look like" rogue water waves. They share the spatial and temporal locality of rogue waves.
In infinite spatial and temporal (both positive and negative) limits, they approach the uniform Stokes waves, and their main humps also have tall enough heights to mimic rogue waves \cite{AAS09}. 

\begin{figure}[ht]
\centering
\includegraphics[width=3.0in,height=1.75in]{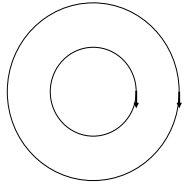}
\caption{Circular orbits on the invariant plane.}
\label{CM}
\end{figure}

One of the simplest deep water weakly nonlinear amplitude model equations is the integrable 1D cubic focusing nonlinear Schr\"odinger equation 
\begin{equation}
iq_t = \pa_x^2 q +2|q|^2q . \label{NLS}
\end{equation}
A simple Peregrine wave solution to (\ref{NLS}) is \cite{DKM08}  \cite{AAS09}
\begin{equation}
q = \left [ 1-4\frac{1-i4t}{1+4x^2+16t^2} \right ] e^{-i2t} . \label{RS} 
\end{equation}
The Peregrine wave solution can be obtained by taking the infinite spatial period limit to the spatially periodic and temporally homoclinic solutions to be discussed below \cite{CS12} \cite{AAS09}.
From now on, we will focus our attention on the Peregrine wave's approximations given by large spatial period homoclinic solutions. Therefore, we pose the spatial periodic boundary condition
\begin{equation}
q(t,x+L ) = q(t,x) \label{BC}
\end{equation}
to (\ref{NLS}). The NLS (\ref{NLS}) with the periodic boundary condition (\ref{BC}) defines a dynamical system in the infinite dimensional phase space $H^1_{[0,L]}$ which is the Sobolev space on the periodic domain $[0, L]$.
Specifically, the norm of $q$ is given by
\[
\| q \|^2_{H^1_{[0,L]}} = \int_0^L ( |q|^2 + |q_x|^2) dx .
\]
One way to visualize dynamics in the infinite dimensional phase space $H^1_{[0,L]}$ is through Fourier series
\[
q(t,x) = \sum_{n\in \mathbb{Z}} q_n(t) e^{inx} ,
\]
where $\mathbb{Z}$ denotes all integers. The set $\{ e^{inx} \}_{n\in \mathbb{Z}}$ forms a base. Each base element $e^{inx}$ spans a complex plane on which the projection of the dynamics is given by $q_n(t)$. In terms of 
$\{  q_n(t) \}_{n\in \mathbb{Z}}$, the NLS (\ref{NLS}) is transformed into infinitely many ordinary differential equations. When $n=0$, the base element $e^{i0x}=1$ spans the spatially independent complex plane $P$ which is a 
two dimensional invaraint subspace under the NLS dynamics. The dynamics on this invariant plane is given by 
\[
iq_t = 2|q|^2q .
\]
The orbits on this  invariant plane are given by the uniform Stokes waves
\begin{equation}
q_c = a e^{-i(2a^2t+\ga )} \label{SW}
\end{equation}
where $a$ is the constant amplitude and $\ga$ is the constant phase. In the original water wave variable, these uniform Stokes waves correspond to the common Stokes water waves. Geometrically, the orbits on the invariant plane are
periodic circular orbits as shown in Figure \ref{CM}.
\begin{figure}[ht]
\centering
\includegraphics[width=3.0in,height=2.5in]{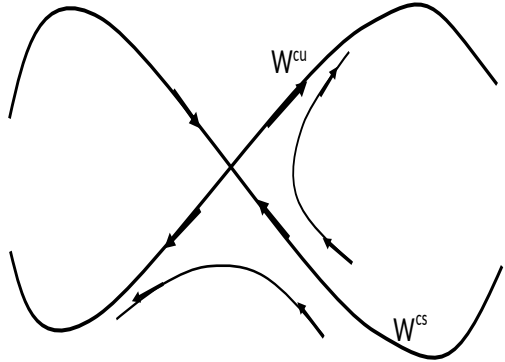}
\caption{An illustration of the Inclination Lemma.}
\label{IL1}
\end{figure}

Observable ocean waves should lie in the neighborhood of the Stokes waves (\ref{SW}) in the infinite dimensional phase space $H^1_{[0,L]}$, and the neighborhood is where we will focus our attention on. Linearize the NLS  (\ref{NLS}) at (\ref{SW})
in the form 
\[
q = a e^{-i(2a^2t+\ga )}(1+Q),
\]
one gets the linearized equation
\[
iQ_t = \pa_x^2 Q + 2a^2 (Q + \bQ ) . 
\]
Set 
\[
Q = A e^{\Om t + i k x } + B e^{\bar{\Om} t - i k x}
\]
where $\Om$, $A$ and $B$ are complex parameters, and $k$ is a real parameter given by 
\[
k = \frac{2\pi}{L} n , \ n \in \mathbb{Z}
\]
to satisfy the boundary condition (\ref{BC}). One gets
\begin{eqnarray*}
&& \left ( [2a^2-k^2]-i \Om \right ) A + 2 a^2 \bar{B} = 0 , \\
&& 2 a^2 A + \left ( [2a^2-k^2]+i \Om \right )\bar{B} = 0 ,
\end{eqnarray*}
and the relation
\[
\Om = \pm \sqrt{[4a^2 - k^2]k^2} .
\]
When 
\begin{equation}
0< k^2< 4a^2, \text{ i.e. } 0<|n|<\frac{aL}{\pi}, \label{MI}
\end{equation}
there is the so-called modulational instability. For any $a>0$, when $L>\frac{\pi}{a}$, the instability appears. That is, no matter how small $a$ is, as long as $L$ is large enough, the instability appears. For fixed $a$ and $L$,
the unstable modes are given by those $n$'s satisfying (\ref{MI}). Let $2N$ be the number of such unstable modes. Then the unstable subspace $S^u$ of the periodic orbit (\ref{SW}) has dimension $2N$, the stable subspace 
$S^s$ of the periodic orbit (\ref{SW}) has dimension $2N$, and the center subspace $S^c$ of the periodic orbit (\ref{SW}) has codimension $4N$. The product of the unstable subspace and the center subspace is the codimension $2N$ 
center-unstable subspace $S^{cu}$, and the product of the stable subspace and the center subspace is the codimension $2N$ center-stable subspace $S^{cs}$. These subspaces can be exponentiated into invariant submanifolds under the 
NLS (\ref{NLS}) dynamics via Darboux transformations \cite{LY14}. We have \cite{Li04}
\begin{theorem} 
Under the NLS (\ref{NLS}) dynamics, the periodic orbit (\ref{SW}) on the invariant plane $P$ has a codimension $4N$ center manifold $W^c$, a codimension $2N$ center-unstable manifold $W^{cu}$, and a 
codimension $2N$ center-stable manifold $W^{cs}$. Moreover, $W^{cu}=W^{cs}$ and $W^{cu}\cap W^{cs}=W^c$.
\end{theorem}
Explicit formulae for certain homoclinic orbits inside $W^{cu}=W^{cs}$ can be found in the Appendix. The neighborhood of the periodic orbit (Stokes wave (\ref{SW})) is divided by  $W^{cu}$ and $W^{cs}$ into different regions. Dynamics 
in the  neighborhood of the periodic orbit follows the following Inclination Lemma \cite{Li03}.
\begin{theorem}  [Inclination Lemma] 
All orbits starting from initial points in the neighborhood of the periodic orbit approach the center-unstable manifold $W^{cu}$ in forward time. 
\end{theorem}
See Figure \ref{IL1} for an illustration. Notice that the center manifold $W^c$ is a measure zero subset of the neighborhood of the periodic orbit, and it is also a measure zero subset of $W^{cu}$. Orbits starting from points inside $W^c$ of course stay inside $W^c$. 
Orbits starting from points inside $W^{cu}$ but not in $W^c$ have the same homoclinic feature as those explicitly calculated in the Appendix. In principle, all such orbits in $W^{cu}$ can be constructed via Darboux transformations 
as shown in the Appendix. One can view all such orbits as rooted to the center manifold $W^c$. In fact, each point in the center manifold $W^c$ is a Fenichel fiber base point, and the Fenichel fibers capture the global features of these 
homoclinic orbits \cite{Li04}. Since the center manifold $W^c$ lies inside the neighborhood of the periodic orbit, those homoclinic orbits rooted to the invariant plane $P$ are good approximations of all such homoclinic orbits 
in $W^{cu}$ which in general may have small amplitude oscillating tails in space and time.  These homoclinic orbits in $W^{cu}$ are generic orbits in $W^{cu}$ in the sense that $W^c$ is a measure zero subset of $W^{cu}$. In view 
of the Inclination Lemma, generic orbits starting from initial points in the neighborhood of the periodic orbit approach those homoclinic orbits in $W^{cu}$ which can be approximated by those homoclinic orbits rooted to the 
invariant plane $P$. The infinite spatial period limits of the homoclinic orbits rooted to the invariant plane $P$ are the Peregrine waves. In conclusion, generic orbits starting from initial points in the neighborhood of the periodic orbit
(Stokes wave) have the homoclinic feature and Peregrine wave feature (when the spatial period approaches infinity). Therefore, such homoclinic orbits and Peregrine waves should be the most observable (common) waves in the deep ocean according to the 
nonlinear Schr\"odinger model. They should not be the rarely observed rogue waves.

When the nonlinear Schr\"odinger equation (\ref{NLS}) is under perturbations (for example by keeping higher order terms in the NLS model of deep water (\ref{PNLS})), the 
center-unstable manifold $W^{cu}$, center-stable manifold $W^{cs}$ and center manifold $W^c$ persist, but $W^{cu}$ and $W^{cs}$ do not coincide anymore \cite{Li04}. Orbits inside $W^{cu}$ have 
a near homoclinic nature. The above conclusion that homoclinic orbits and Peregrine waves should be the most observable common waves rather than rogue waves, still holds.

\section{Conclusion and discussion}

Based upon the above rigorous mathematical analysis on the infinite dimensional phase space where the nonlinear Schr\"odinger equation (\ref{NLS}) defines a dynamical system, we conclude that Peregrine waves and 
homoclinic orbits are the waves most commonly observable in deep ocean rather than rogue water waves. Next we discuss two other possibilities for the mechanism of rogue waters.

\subsection{Rough dependence on initial data}

The solution operator of high Reynolds number Navier-Stokes equations has rough dependence on initial data \cite{Li14} \cite{Li15}. Temporal amplification of certain perturbations to the initial data can potentially reach
\begin{equation}
\sim \ e^{\sg \sqrt{Re} \sqrt{t}} , \label{est}
\end{equation}
where $\sg$ is a constant and $Re$ is the Reynolds number. When the Reynolds number is large, such amplification can reach substantial amount in very short time. This feature of the solution operator may explain the
(no apparent reason) sudden amplification of one wave among many into a rogue wave in the deep ocean \cite{Cha09}. That particular wave may receive just the right perturbation which amplifies superfast like the above 
estimate, and very quickly develops into a rogue wave. In this sense, the choice of the particular wave is random, the right perturbation is random, and the temporal and spatial locations of the event are also random. All 
these factors may manifest into a sudden appearance of a rogue wave. High Reynolds number Navier-Stokes equations are good models of water waves since real fluids 
(water or air) always have viscosity (no matter how slight it may be). On the other hand, for simplicity, most mathematical models of water waves are derived from Euler equations, and the solution operator of the Euler equations is 
nowhere differentiable in its initial data \cite{Inc15} (formally one can set $Re$ to infinity in the above estimate (\ref{est})) . 

\subsection{Finite time blowup}

A great open problem is whether or not water wave equations have finite time blowup solutions. A 
hint of finite time blowup solutions comes from simple nonlinear wave equations, for example, the 
one dimensional nonlinear Schr\"odinger equation
\begin{equation}
iq_t = \pa_x^2q+|q|^{s-1} q, \label{SNLS}
\end{equation}
where $q(t,x)$ is a complex-valued function of ($t,x$). For the initial condition of the form
\[
q(0,x) = e^{ix^2} \psi (x),
\]
where $\psi (x)$ is a real-valued function, when the initial energy
\[
\int_{\mathbb{R}} \bigg ( |\pa_x q(0,x)|^2- \frac{2}{s+1} |q(0,x)|^{s+1} \bigg ) dx 
\]
is non-positive and $s\geq 5$, the solution blows up in finite time \cite{Gla77} \cite{Mer92} 
\cite{Mer93} \cite{Bou99}. That is, there is a finite time $0<T<\infty$, such that 
\[
\lim_{t \ra T^-} \| q(t,x) \|_{L^\infty} = \infty, \ \ \lim_{t \ra T^-} \| \pa_x q(t,x) \|_{L^2} = \infty .
\]
Such a finite time blowup solution resembles very much a rogue wave in terms of spatially and temporally 
local nature. One should only take such a finite time blowup solution as a hint rather than a clear indication for a possible finite time blowup solution to the 
water wave equations. There are a lot of simple models of water wave equations, for example, the Davey-Stewartson equations \cite{DS74}. For the Davey-Stewartson equations with coefficients in the water 
wave regime, a finite time blowup solution has not been found. For the Davey-Stewartson equations with coefficients outside the water wave regime, finite time 
blowup solutions have been found \cite{Oza92}. In the deep water limit, the Davey-Stewartson 
equations \cite{DS74} reduce to the following equation
\begin{equation}
iq_t = \square q + 2 |q|^2q \label{HNLS}
\end{equation}
where $q(t,x,y)$ is complex-valued and 
\[
\square = \pa_x^2 - \pa_y^2 .
\]
This equation has two conserved quantities
\[
I = \int |q|^2 dx dy , 
\]
\[
E = \int [ |\pa_xq|^2 - |\pa_yq|^2 - |q|^4 ] dxdy . 
\]
Since the two conserved quantities do not bound $H^1$ norm, this equation may have finite time blowup solutions. When the operator $\square$ is replaced by 
\[
\Dl = \pa_x^2 +\pa_y^2 ,
\]
there are indeed finite time blowup solutions \cite{Bou99}. Linearize equation (\ref{HNLS}) at 
\[
q_* = a e^{-i(2a^2t+\th )} 
\]
where $a > 0$ is the amplitude and $\th$ is the phase, in the form 
\[
q = a e^{-i(2a^2t+\th )}(1+Q),
\]
one gets the linearized equation
\[
iQ_t = \square Q + 2a^2 (Q + \bQ ) . 
\]
Set 
\[
Q = A e^{\Om t + i k_1 x +ik_2 y} + B e^{\bar{\Om} t - i k_1 x -ik_2 y}
\]
where $\Om$, $A$ and $B$ are complex parameters, and ($k_1,k_2$) are real parameters, one gets
\begin{eqnarray*}
&& \left ( [(k_2^2-k_1^2)+2a^2]-i \Om \right ) A + 2 a^2 \bar{B} = 0 , \\
&& 2 a^2 A + \left ( [(k_2^2-k_1^2)+2a^2]+i \Om \right )\bar{B} = 0 ,
\end{eqnarray*}
and the relation
\[
\Om = \pm \sqrt{[4a^2 - (k_1^2-k_2^2)](k_1^2-k_2^2)} .
\]
When 
\[
0< k_1^2-k_2^2 < 4a^2, 
\]
there is a modulational instability. 

In one spatial dimension, equation (\ref{HNLS}) reduces to the integrable cubic nonlinear Schr\"odinger equation (\ref{NLS}).
By keeping higher order terms, the one spatial dimension
deep water wave model can be written as 
\begin{equation}
iq_t = \pa_x^2 q + 2 |q|^2q + H(q) \label{PNLS}
\end{equation}
where $H(q)$ represents the higher order terms which may involve a variety of terms like higher order derivatives and higher order nonlinearities \cite{Dys79}. With the higher order terms in, equation (\ref{PNLS})
may have finite time blowup solutions. Invoking possible finite time blowup solutions to models of water wave equations is paradoxical in the search for finite time blowup solutions to the full water wave equations. 
Most of these models are derived under the assumption of weak nonlinearity, while finite time blowup is a strongly nonlinear phenomenon. 

\section{Appendix: Explicit formulae of homoclinic orbits}

\begin{figure}[ht]
\centering
\includegraphics[width=4.5in,height=1.0in]{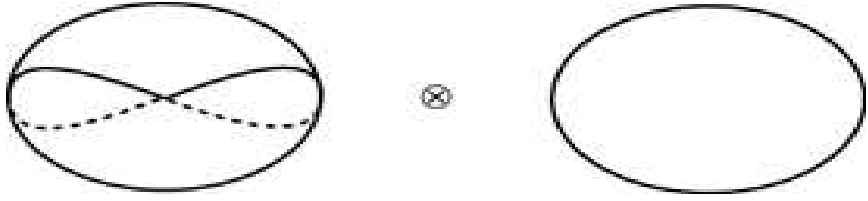}
\caption{Figure eight structure of noneven data with one unstable mode.}
\label{snef}
\end{figure}

Let $L = 2\pi$. When 
\[
\frac{1}{2} < a <1 ,
\]
the Stokes wave (\ref{SW}) has one linearly unstable mode, and when 
\[
1 < a < \frac{3}{2},
\]
the Stokes wave (\ref{SW}) has two linearly unstable modes, etc. The homoclinic orbits asymptotic to the 
Stokes wave (\ref{SW}) are the nonlinear amplifications of the linearly unstable modes. When 
$\frac{1}{2} < a <1$, the homoclinic orbit is given by \cite{Li04a}
\begin{eqnarray}
q_1 &=& q_c \bigg [ 1 + \sin \vth_0 \ \mbox{sech} \tau 
\cos y \bigg ]^{-1} \cdot \bigg [ \cos 2\vth_0 - i \sin 2\vth_0 \tanh \tau
 \nonumber \\
& & - \sin \vth_0 \ \mbox{sech} \tau \cos y \bigg ] \ , \label{sne}
\end{eqnarray}
where
\begin{equation}
\tau = 2 \sg t - \rho \ , \ \ y = x + \vth - \vth_0 +\pi/2\ ,
\label{par1}
\end{equation}
where $\sg$, $\rho$, $\vth$ and $\vth_0$ are real parameters.
As $t \ra \pm \infty$, 
\begin{equation}
q_1 \ra q_c e^{\mp i2\vth_0}\ .
\label{asy1}
\end{equation}
Thus $q_1$ is asymptotic to $q_c$ up to phase shifts as $t \ra \pm \infty$.
We say $Q$ is a homoclinic orbit asymptotic to the periodic orbit given 
by $q_c$. For a fixed amplitude $a$ of $q_c$, the phase $\ga$ of $q_c$ and 
the B\"acklund parameters $\rho$ and $\vth$ parametrize a $3$-dimensional 
submanifold with a figure eight structure. For an illustration, see 
Figure \ref{snef}. 
\begin{figure}[ht]
\centering
\includegraphics[width=4.5in,height=1.0in]{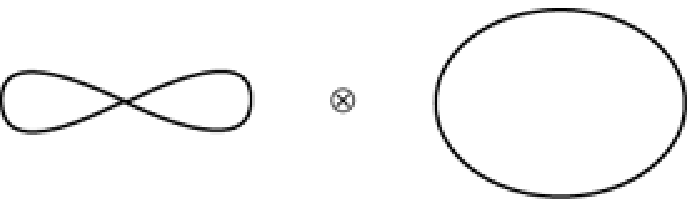}
\caption{Figure eight structure of even data with one unstable mode.}
\label{sef}
\end{figure}
\nid 
If one restricts the 
B\"acklund parameter $\vth$ by $\vth -\vth_0 +\pi/2 = 0$, or $\pi$, 
one gets $q_1$ to be even in $x$, 
\begin{eqnarray}
q_1 &=& q_c \bigg [ 1 \pm \sin \vth_0 \ \mbox{sech} \tau  
\cos x \bigg ]^{-1} \nonumber \\
& & \cdot \bigg [ \cos 2\vth_0 - i \sin 2\vth_0 \tanh \tau \mp 
\sin \vth_0 \ \mbox{sech} \tau \cos x \bigg ] \ , \label{se}
\end{eqnarray}
where the upper sign corresponds to $\vth -\vth_0 +\pi/2 =0$. Then for a fixed amplitude 
$a$ of $q_c$, the phase $\ga$ of $q_c$ and 
the B\"acklund parameter $\rho$ parametrize a $2$-dimensional 
submanifold with a figure eight structure. For an illustration, see 
Figure \ref{sef}. 

\begin{figure}[ht]
\centering
\includegraphics[width=4.5in,height=1.0in]{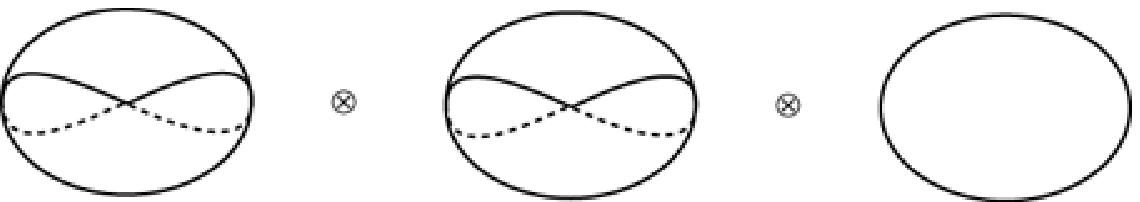}
\caption{Figure eight structure of noneven data with two unstable modes.}
\label{dnef}
\end{figure}

When $1< a <\frac{3}{2}$, the homoclinic orbit is given by \cite{Li04a}
\begin{equation}
q_2 = q_1 + q_c \frac{\W_2 \sin \hvth_0}{\W_1}\ ,
\label{dne}
\end{equation}
where $q_1$ is given in (\ref{sne}),
\begin{eqnarray*}
\W_1 &=& \bigg [ (\sin \hvth_0)^2(1+\sin \vth_0 \ \mbox{sech} \tau 
\cos y )^2 +\frac{1}{8} (\sin 2\vth_0)^2 (\mbox{sech} \tau)^2 
(1 -\cos 2y) \bigg ] \\
& & \cdot (1 + \sin \hvth_0 \ \mbox{sech} \htau 
\cos \hy ) \\
&-& \frac {1}{2} \sin 2\vth_0 \sin 2\hvth_0 \ \mbox{sech} \tau 
\ \mbox{sech} \htau (1+\sin \vth_0 \ \mbox{sech} \tau 
\cos y ) \sin y \sin \hy  \\
&+& (\sin \vth_0)^2 \bigg [ 1+ 2 \sin \vth_0 \ \mbox{sech} \tau 
\cos y + [(\cos y)^2 - (\cos \vth_0)^2](\mbox{sech} \tau)^2 \bigg ]
\\
& &\cdot (1 + \sin \hvth_0 \ \mbox{sech} \htau 
\cos \hy ) \\
&-& 2\sin \hvth_0 \sin \vth_0 \bigg [ \cos \hvth_0 \cos \vth_0
\tanh \htau \tanh \tau + ( \sin \vth_0 + \ \mbox{sech} \tau 
\cos y)\\
& & \cdot ( \sin \hvth_0 + \ \mbox{sech} \htau 
\cos \hy) \bigg ] (1 + \sin \vth_0 \ \mbox{sech} \tau 
\cos y ) \ ,
\end{eqnarray*}
\begin{eqnarray*}
\W_2 &=& \bigg [ -2 (\sin \hvth_0)^2 (1 + \sin \vth_0 \ 
\mbox{sech} \tau \cos y )^2 +\frac {1}{4} (\sin 2\vth_0)^2 
(\mbox{sech} \tau)^2 (1-\cos 2y)\bigg ]\\
& & \cdot (\sin \hvth_0 + 
\ \mbox{sech} \htau \cos \hy + i \cos \hvth_0 \tanh \htau ) \\
&+& 2 (\sin \vth_0)^2(-\cos \vth_0 \tanh \tau + i \sin \vth_0 + 
i\ \mbox{sech} \tau \cos y)^2 \\
& & \cdot (\sin \hvth_0 + 
\ \mbox{sech} \htau \cos \hy - i \cos \hvth_0 \tanh \htau )  \\
&+& 2 \sin \vth_0 (\sin \vth_0 + 
\ \mbox{sech} \tau \cos y + i \cos \vth_0 \tanh \tau ) 
\\
& & \cdot \bigg [2 \sin \hvth_0 (1 + \sin \vth_0 \ 
\mbox{sech} \tau \cos y )(1 + \sin \hvth_0 \ 
\mbox{sech} \htau \cos \hy )\\
& & - \sin 2\vth_0 \cos \hvth_0 
\ \mbox{sech} \tau \ \mbox{sech} \htau  \sin y \sin \hy \bigg ] \ ,
\end{eqnarray*}
where some of the notations are given in (\ref{sne}), and
\[
\hat{\tau} = 4 \hat{\sg} t - \hat{\rho} \ ,
\ \ \hy = 2x + \hvth - \hvth_0 +\pi/2\ ,
\]
and $\hat{\sg}$, $\hat{\rho}$, $\hvth$ and $\hvth_0$ are real parameters.
The asymptotic phase of $q_2$ is as follows, as $t \ra \pm \infty$,
\begin{equation}
q_2 \ra q_c e^{\mp i 2 (\vth_0 + \hvth_0)}\ . 
\label{dasym}
\end{equation}
Thus $q_2$ is asymptotic to $q_c$ up to phase shifts as 
$t \ra \pm \infty$. For a fixed amplitude 
$a$ of $q_c$, the phase $\ga$ of $q_c$ and the B\"acklund parameters $\rho$, $\vth$, $\hat{\rho}$, and $\hvth$ parametrize a $5$-dimensional 
submanifold with a figure eight structure. For an illustration, see 
Figure \ref{dnef}. 
\begin{figure}[ht]
\centering
\includegraphics[width=4.5in,height=1.0in]{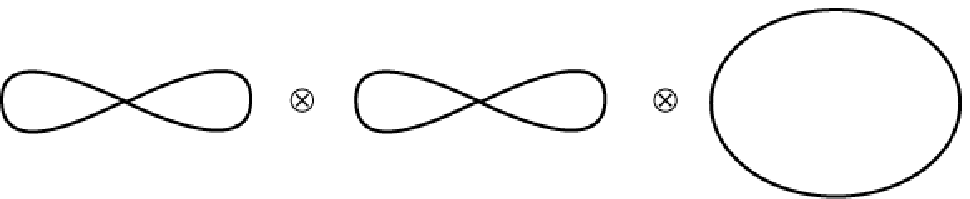}
\caption{Figure eight structure of even data with two unstable modes.}
\label{def}
\end{figure}
If one put restrictions on the B\"acklund parameters $\vth$ and 
$\hvth$, s.t. 
\begin{equation}
\vth - \vth_0 +\pi/2 = \left \{ \begin{array}{c} 0 
\ \left \{ \begin{array}{c} \hvth - \hvth_0 +\pi/2 = 0 \ ,\cr
\hvth - \hvth_0 +\pi/2 = \pi \ , \cr \end{array} \right.
\cr \pi \ \left \{ \begin{array}{c} \hvth - \hvth_0 +\pi/2 = 0 \ ,\cr
\hvth - \hvth_0 +\pi/2 = \pi \ , \cr \end{array} \right.
\cr \end{array} \right.
\label{dec}
\end{equation}
then $q_2$ is even in $x$. Thus for a fixed amplitude 
$a$ of $q_c$, the phase $\ga$ of $q_c$ and the B\"acklund parameters $\rho$ and $\hat{\rho}$ parametrize a $3$-dimensional 
submanifold with a figure eight structure. For an illustration, see Figure \ref{def}.


\begin{thebibliography}{99}

\bibitem{AAS09} N. Akhmediev, A. Ankiewicz, J. Soto-Crespo, Rogue waves and rational solutions of the 
nonlinear Schr\"odinger equation, {\it Phys. Rev. E} {\bf 80} (2009), 026601.

\bibitem{Ala14} M. Alam, Predictability horizon of oceanic rogue waves, {\it Geophysical Research Letters} {\bf 41, no.23} (2014), 8477-8485.

\bibitem{Bou99} J. Bourgain, {\it Global Solutions of Nonlinear Schr\"odinger Equations}, AMS 
Colloquium Publications, vol.46, 1999.

\bibitem{CS12} A. Calini, C. Schober, Dynamical criteria for rogue waves in nonlinear Schr\"odinger models,
{\it Nonlinearity} {\bf 25} (2012), R99-R116.

\bibitem{CHA11} A.  Chabchoub, N. Hoffmann, N. Akhmediev, Rogue wave observation in a water wave tank, {\it Phys. Rev. E} {\bf 106} (2011), 204502.

\bibitem{Cha09} D. Chalikov, Freak waves: Their occurrence and probability, {\it Physics of Fluids} {\bf 21} (2009), 076602.

\bibitem{DS74} A. Davey, K. Stewartson, On three-dimensional packets of surface waves, {\it Proc. R. 
Soc. Lond. A} {\bf 338} (1974), 101-110.

\bibitem{Dys79} K. Dysthe, Note on a modification to the nonlinear Schr\"odinger equation for deep water, 
{\it Proc. R. Soc. Lond. A} {\bf 298} (1979), 105-114.

\bibitem{DKM08} K. Dysthe, H. Krogstad, P. M\"uller, Oceanic rogue waves, {\it Ann. Rev. Fluid Mech.} {\bf 40} (2008), 287-310.

\bibitem{DT99} K. Dysthe, K. Trulsen, Note on breather type solutions of the NLS as model for freak waves,
{\it Phys. Scr.} {\bf T82} (1999), 48-52. 

\bibitem{Gla77} R. Glassey, On the blowing up of solutions to the Cauchy problem for nonlinear Schr\"odinger equations, {\it J. Math. Phys.} {\bf 18, no.9} (1977), 1794-1797.

\bibitem{GZT13} O. Gramstad, H. Zeng, K. Trulsen, G. Pedersen, Freak waves in weakly nonlinear unidirectional wave trains over a sloping bottom in shallow water, {\it Physics of Fluids} {\bf 25} (2013), 122103.

\bibitem{Inc15} H. Inci, On the regularity of the solution map of the incompressible Euler equation, {\bf Dynamics of PDE} {\bf 12, no.2} (2015), 97-113.

\bibitem{KP03} C. Kharif, E. Pelinovsky, Physical mechanisms of the rogue wave phenomenon, {\it European J. Mech. - B/Fluids} {\bf 22, no.6} (2003), 603-634.

\bibitem{Li03} Y. Li, Chaos and shadowing lemma for autonomous systems
of infinite dimensions, {\it J. Dyn. Diff. Eq.} {\bf 15, no.4} (2003), 699-730. 

\bibitem{Li04} Y. Li, {\it Chaos in Partial Differential Equations},
International Press, (2004). 

\bibitem{Li04a} Y. Li,  Persistent homoclinic orbits for nonlinear Schr\"odinger equation
under singular perturbation, {\it Dynamics of PDE} {\bf 1, no.1} (2004), 87-123.

\bibitem{Li14} Y. Li, The distinction of turbulence from chaos --- rough dependence on initial data, {\it Electronic Journal of Differential Equations} {\bf 2014, no. 104}  (2014), 1-8.

\bibitem{Li15} Y. Li, Rough dependence upon initial data exemplified by explicit solutions and the effect of viscosity, {\it  arXiv:1506.05498}.

\bibitem{LY14} Y. Li, A. Yurov, {\it Lie-B\"acklund-Darboux Transformations},
International Press, Boston, USA and Higher Education Press, Beijing, China
Surveys of Modern Mathematics, Vol.8, (2014).
 
\bibitem{Mer92} F. Merle, Construction of solutions with exact k blow-up points for the Schr\"odinger equation with critical power nonlinearity, {\it Comm. Math. Phys.} {\bf 149} (1992), 205-214.

\bibitem{Mer93} F. Merle, Determination of blow-up solutions with minimal mass for nonlinear Schr\"odinger equation with critical power, {\it Duke Math. J.} {\bf 69} (1993), 427-453.

\bibitem{Oza92} T. Ozawa, Exact blow-up solutions to the Cauchy problem for the Davey-Stewartson systems, {\it Proc. R. Soc. Lond. A} {\bf 436} (1992), 345-349. 

\bibitem{SPS13} A. Slunyaev, E. Pelinovsky, A. Sergeeva, A.  Chabchoub, N. Hoffmann, M. Onorato, N. Akhmediev, Super-rogue waves in simulations based on weakly nonlinear and fully nonlinear hydrodynamic 
equations, {\it Phys. Rev. E} {\bf 88} (2013), 012909.


\end{thebibliography}
\end{document}